\documentclass[prb,showpacs,twocolumn,preprintnumbers,amsmath,amssymb,floatfix]{revtex4}
\usepackage[dvips]{graphicx}
\usepackage[latin1]{inputenc}
\begin{document}
\draft

\newcommand{\liti} {Li$_2$V$_{1-x}$Ti$_x$OSiO$_4$}
\newcommand{\lisi} {Li$_2$VOSiO$_4$}
\newcommand{\etal} {{\it et al.} }
\newcommand{\ie} {{\it i.e.} }
\newcommand{\aucr}{CeCu$_{5.9}$Au$_{0.1}$ }
\newcommand{\auaf}{CeCu$_{5.2}$Au$_{0.8}$ }
\newcommand{\aux}{CeCu$_{6-x}$Au$_{x}$ }
\newcommand{\ip}{${\cal A}^2$ }

\hyphenation{a-long}

\title{SPIN DILUTION IN FRUSTRATED
TWO-DIMENSIONAL S=1/2 ANTIFERROMAGNETS ON A SQUARE LATTICE }

\author{N. Papinutto, and P. Carretta}

\affiliation{Department of Physics ``A.Volta", University of
Pavia, Via Bassi 6, I-27100, Pavia (Italy)}

\author{S. Gonthier and P. Millet}

\affiliation{CEMES, CNRS, 31055 Toulouse Cedex, France}

\widetext

\begin{abstract}

$^7$Li and $^{29}$Si NMR, $\mu$SR and magnetization measurements
in Li$_2$V$_{1-x}$OTi$_x$SiO$_4$, for $0 \leq x \leq 0.2$, are
presented. The $x=0$ compound is a prototype of frustrated
two-dimensional Heisenberg antiferromagnet on a square-lattice
with competing nearest ($J_1$) and next-nearest ($J_2$) neighbour
exchange interactions. Ti$^{4+}$ (S=0) for V$^{4+}$ (S=1/2)
substitution yields the spin dilution of the antiferromagnetic
layers. The analysis of the magnetization and of the nuclear
spin-lattice relaxation rate shows that spin dilution not only
reduces the spin-stiffness by a factor $\simeq (1-x)^2$, but also
causes the decrease of the effective ratio $J_2(x)/J_1(x)$.
Moreover, the sublattice magnetization curves derived from
zero-field $\mu$SR measurements in the collinear phase point out
that, at variance with non-frustrated two-dimensional Heisenberg
antiferromagnets, spin dilution affects the low-temperature
staggered magnetization only to a minor extent. This observation
is supported also by the $x$ dependence of the collinear ordering
temperature. The results obtained for the Ti doped samples are
discussed in the light of the results previously obtained in the
pure $x=0$ compound and in non-frustrated two-dimensional
Heisenberg antiferromagnets with spin-dilution.

\end{abstract}

\pacs {76.60.Es, 76.75.+i, 75.10.Jm} \maketitle

\narrowtext

\section{Introduction}

The search for novel quantum states in low-dimensional
antiferromagnets (AF) has triggered a significant activity in
recent times \cite{Rice}. A remarkable amount of theoretical
studies has concerned the phase diagram of antiferromagnets where
long-range magnetic order is suppressed by enhanced quantum
fluctuations. Such a scenario can be established when the magnetic
lattice dimensionality and the spin value are reduced or when the
disorder is increased, either by means of heterovalent
substitutions or by spin dilution \cite{CHN,Rep,NS}. Further
enhancement of quantum fluctuations can occur when
antiferromagnetic interactions are competing as, for instance, in
frustrated two-dimensional $S=1/2$ Heisenberg AF (2DQHAF) on a
square-lattice, with nearest neighbour ($J_1$) and next-nearest
neighbour ($J_2$) antiferromagnetic couplings  of the same order
of magnitude \cite{Chandra}.

Recently, a prototype of frustrated 2DQHAF on a square-lattice has
been discovered, the \lisi \cite{Melzi1}. The analysis of the
magnetic susceptibility, of the specific heat and of other
quantities \cite{Melzi2} indicate that this compound is
characterized by a ratio $J_2/J_1$ ranging from 1 to 4 \cite{JPC}.
The ground state is collinear \cite{Melzi1,Bombardi}, as
theoretically expected \cite{Chandra2}, and the staggered
magnetization reaches a value $\simeq 0.6 \mu_B/$V$^{4+}$
\cite{Bombardi}, consistent with the $J_2/J_1$ estimate
\cite{Schulz}. This value is remarkably close to the one of a
non-frustrated 2DQHAF \cite{Keimer} and, therefore, at first sight
one could think that, as far as $J_2/J_1$ is not close to $0.5$
and quantum fluctuations are not so strong, frustration does not
affect sizeably the static magnetic properties of a 2DQHAF. In
order to unravel the basic differences in the properties of
frustrated and non-frustrated 2DQHAF it is tempting to compare
also the behaviour of these systems when disorder is introduced,
for example, by spin dilution. Spin dilution has been widely
investigated in prototypes of non-frustrated 2DQHAF, as
La$_2$Cu$_{1-x}$(Zn,Mg)$_x$O$_4$ \cite{Corti,Sala,NS,Neto}, and
evidence for the validity of the simple dilution model with a
renormalization of the spin stiffness and for the disappearance of
long-range magnetic order at the classical percolation threshold
has been given \cite{NS,Sala}. At first, in a frustrated 2DQHAF on
a square-lattice one would expect that the enhancement of quantum
fluctuations  leads to a suppression of the long-range order well
below the percolation threshold. However, one should also consider
that, owing to the next-nearest neighbours coupling, the
percolation threshold extends to much larger doping levels with
respect to non-frustrated antiferromagnets \cite{Tommaso}. Hence,
it is not trivial to say a priori how the magnetic properties vary
upon increasing the spin dilution.

In the following an experimental study of spin dilution effects in
frustrated 2DQHAF on a square-lattice will be presented. In \lisi
\, spin dilution can be achieved upon Ti$^{4+}$ ($S=0$) for
V$^{4+}$ ($S=1/2$) substitution. It is found that spin dilution
not only reduces the spin-stiffness by a factor $\simeq (1-x)^2$,
but also causes a reduction of the effective ratio
$J_2(x)/J_1(x)$. The analysis of the sublattice magnetization
curves shows that, at variance with non-frustrated 2DQHAF, spin
dilution affects the low-temperature staggered magnetization only
to a minor extent. The $x$-dependence of the collinear ordering
temperature and of other quantities will be discussed vis-\'a-vis
with the trend observed for non-frustrated 2DQHAF. In the
following section a description of the technical aspects involved
in the sample preparation, magnetization, NMR and $\mu$SR
measurements will be presented together with the experimental
data. Then the results will be discussed in the light of models
which turned out to be valid for non-frustrated spin-diluted
2DQHAF. The final conclusions will be given in Sect.IV.

\section{Technical Aspects and Experimental results}

\subsection{Sample preparation and Magnetization measurements}

Powder samples of \liti \, were prepared in a platinum crucible by
solid state reaction starting from a stoichiometric mixture of
Li$_2$SiO$_3$ (Aldrich, 99.5\% ), TiO$_2$ (Aldrich, 99.8\% ) and
VO$_2$ at 830 C, under vacuum, for 24h \cite{Millet}. VO$_2$ was
prepared from a stoichiometric mixture of V$_2$O$_5$ and
V$_2$O$_3$ by heating in a vacuum-sealed quartz tube at 650 C for
24h. V$_2$O$_3$ itself was obtained by reducing V$_2$O$_5$
(99.9\%, Aldrich Chem. Co.) under hydrogen at 800 C. The sample
purity was analyzed by means of X-ray powder diffraction and all
peaks corresponded to the ones of \lisi . The lattice parameters
of each sample were refined using the progamm CELREF. The
substitution of Ti for V leads to no significative variation of
the cell parameters, which is consistent with the rather similar
ionic radii of the two ions ($r_V = 0.53$ \AA and $r_{Ti}= 0.51$
\AA \cite{XRD}). Upon varying $x$ from 0 to 0.15 the $a$ axes
varies from 6.3683(4) \AA \, to 6.3678(6) \AA \, while the $c$
axes varies from 4.449(3) \AA \, to 4.4502(4) \AA . The
homogeneity of Ti concentration in the different samples was
verified by EDX (Energy Dispersive X-ray Analysis).

\begin{figure}[b!]
\vspace{6cm} \includegraphics{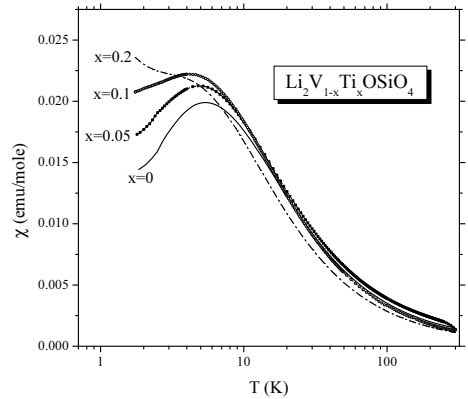} \caption{Temperature dependence of the
susceptibility in Li$_2$V$_{1-x}$OTi$_x$SiO$_4$ for different Ti
contents, in a magnetic field of 1 kGauss.}\label{figsquid}
\end{figure}

Magnetization measurements on \liti \, powders were performed
using a Quantum Design XPMS-XL7 SQUID magnetometer. The
temperature dependence of the susceptibility, defined as $\chi=
M/H$ is shown in Fig.\ref{figsquid}. One observes a high
temperature Curie-Weiss behaviour, a low-temperature maximum
indicating the onset of antiferromagnetic correlations and a small
kink in the 2-3 K range signaling a phase transition to a
collinear ground-state. In the more doped sample, with 20\% of Ti,
a low-temperature upturn is noticed. This upturn is typical of
diluted 2DQHAF \cite{chiZn} and originates from the correlated
response of the spins around the $S=0$ impurity. Above 20 K the
data can be conveniently fitted according to
\begin{equation}
\chi= \frac{C}{T+\Theta} + \chi_{VV} ,
\end{equation}
where $C$ is Curie constant, $\Theta$ the Curie-Weiss temperature
and $\chi_{VV}=4\times 10^{-4}$ emu/mole \cite{Melzi1} the
T-independent Van-Vleck susceptibility \cite{Melzi1}. The
x-dependence of the Curie-Weiss temperature is shown in Fig.2,
together with the x-dependence of the temperature $T^m_\chi$ at
which $\chi$ displays a maximum and of the transition temperature
$T_c$ to the collinear phase. This latter quantity was estimated
from the peak in the derivative $d\chi/dT$. One notices basically
a monotonous decrease of all three quantities with increasing
dilution, as expected.

\begin{figure}[b!]
\vspace{9cm} \includegraphics{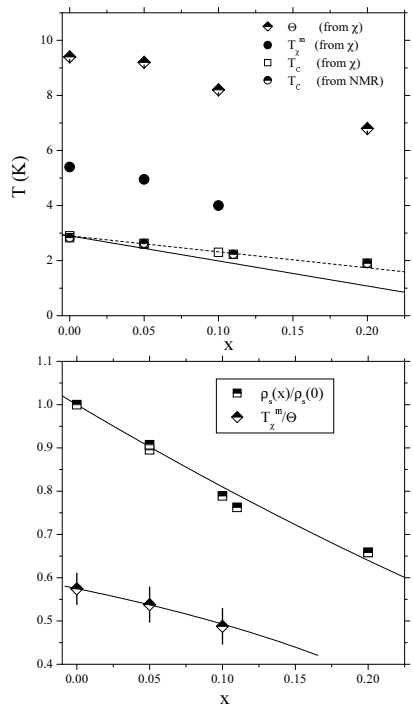} \caption{(Top) Doping dependence of
$T_{\chi}^m$, $\Theta$ and $T_c$ as derived from susceptibility
and NMR measurements. Lines track the initial suppression relation
$-d\,T_c(x)/d\,x=C\,T_c(0)$ described in the text, with $C=3.2$
(solid line) and $C=2$ (dotted line). (Bottom) The spin stiffness
$\rho_s(x)$, derived from Eq. 5 (see text), and the ratio
$T_{\chi}^m(x)/\Theta(x)$ are reported for different Ti contents.
The solid line passing through $\rho_s(x)/\rho_s(0)$ data is the
function $(1-x)^2$. The one through $T_{\chi}^m(x)/\Theta(x)$
points is a guide to the eye. }\label{fig:4}
\end{figure}

\subsection{NMR measurements}

\begin{figure}[b!]
\vspace{6cm} \includegraphics{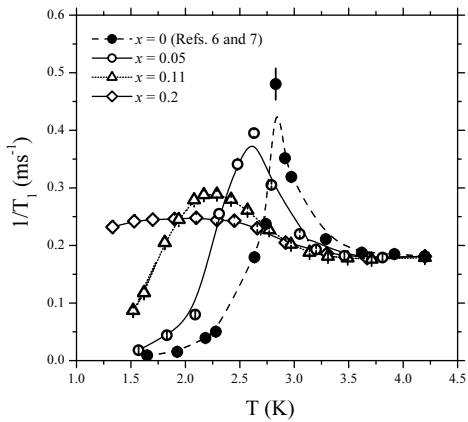} \caption{Temperature dependence of $^7$Li
nuclear spin-lattice relaxation rate in
Li$_2$V$_{1-x}$OTi$_x$SiO$_4$ powders for different doping amounts
$x$. The $x=0$ data \cite{Melzi1,Melzi2} were measured on a single
crystal with $\vec H \parallel c$. In order to compare them with
the powder sample data they were divided by a factor $1.18$ owing
to the different hyperfine coupling involved. The lines are guide
to the eye. }\label{fig:2}
\end{figure}
\begin{figure}[b!]
\vspace{6cm} \includegraphics{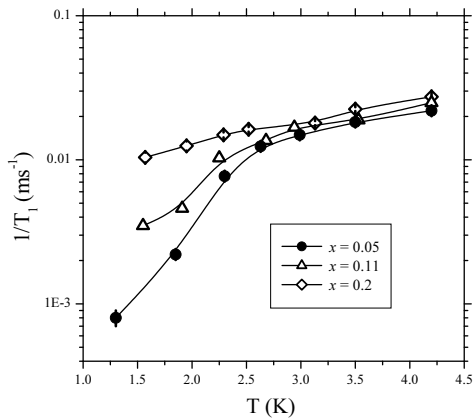} \caption{Temperature dependence of
$^{29}$Si nuclear spin-lattice relaxation rate in
Li$_2$V$_{1-x}$OTi$_x$SiO$_4$ for different doping amounts $x$.
The lines are guide to the eye. }\label{fig:3}
\end{figure}

$^7$Li and $^{29}$Si NMR measurements have been carried out using
standard radiofrequency pulse sequences. In particular, the
nuclear spin-lattice relaxation rate $1/T_1$  was estimated from
the recovery of the nuclear magnetization after a saturating pulse
sequence. Since in the powder sample at low-T $^7$Li NMR line
broadens due to the paramagnetic shift anisotropy, only a partial
irradiation of the central and satellite lines is achieved. This
causes a departure of the recovery law from a single exponential
and hence at low temperature, below about 3 K, it is more
appropriate to fit the recovery of the nuclear magnetization with
a stretched exponential form, namely
$M(\tau)=M(\infty)\cdot(1-exp\,(-(\tau/T_1)^\beta))$. The exponent
$\beta$ decreased down to $\beta\simeq 0.7$ for $T\simeq 1.3$ K.
On the other hand, the recovery law for $^{29}$Si was a single
exponential over all the explored T-range. The temperature
dependence of $^7$Li and $^{29}$Si $1/T_1$ derived from the fit of
the recovery laws following the aforementioned procedure are shown
in Figs. 3 and 4, respectively. One notices that while $^7$Li
$1/T_1$ displays a peak at $T_c$, which broadens upon increasing
$x$, $^{29}$Si $1/T_1$ doesn't show any peak and the relaxation
rate is found to decrease on cooling.

\subsection{$\mu$SR measurements}

$\mu$SR measurements were performed on the EMU instrument at the
ISIS pulsed muon facility, using $29$ MeV/c spin-polarized muons.

{\liti} powders were pressed on a silver sample-holder, whose
background contribution to the muon asymmetry was determined from
the slowly decaying oscillating signal in a 100 Gauss transverse
magnetic field, below $T_c$. In fact, below $T_c$ the external
magnetic field sums up with the randomly oriented internal field
$H_{int}$ and gives rise to a fast damping of the oscillating
signal due to the muons stopping in the sample. The background was
estimated $A_{back}\simeq 0.071$ for $x=0.05$ and
$A_{back}=0.0475$ for $x=0.11$. These values were kept fixed for
all the subsequent fits.

\begin{figure}[b!]
\vspace{9cm} \includegraphics{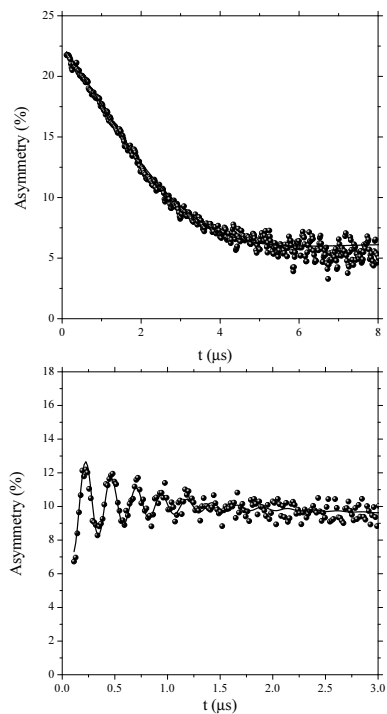}  \caption{$\mu$SR asymmetry in
Li$_2$V$_{1-x}$OTi$_x$SiO$_4$ for $x=0.05$ at $2.8$\,K (top) and
$0.5\,$K (bottom). The solid line in the lower plot is the fit
according to Eq.\,(\ref{muons2}) in the text, while the one in the
upper plot is the fit according to Eq.3.}\label{muons1}
\end{figure}

\begin{figure}[t!]
\vspace{4.8cm} \includegraphics{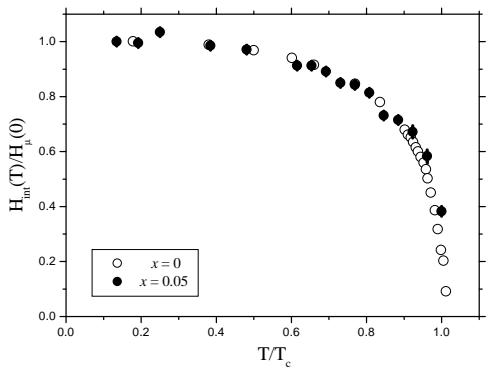} \caption{Temperature dependence of the
local field at the muon site $H_{int}$ in
Li$_2$V$_{1-x}$OTi$_x$SiO$_4$ for $x=0.05$, obtained from
Eq.\,(\ref{muons2}) in the text. Data for Li$_2$VOSiO$_4$ (from
Ref.\,\onlinecite{musr}) are reported for comparison. Fields and
temperatures are normalized by $H_{int}(0)=309.2\,G$ and
$T_c=2.6\,K$ for $x=0.05$ and $H_{int}(0)=313\,G$ and
$T_c=2.86\,K$ for $x=0$. For the $x=0.11$ a value for
$H_{int}(0)=297 \pm 5\,G$ was derived.}\label{field}
\end{figure}

By means of zero field (ZF) muon spin relaxation measurements it
is possible to extract the temperature dependence of the order
parameter below $T_c$, as previously done for the pure
Li$_2$VOSiO$_4$ \cite{musr}. Below $T_c$, superposed to an almost
constant background, one observes a precessional signal at
frequency $\omega_{\mu}=\gamma_{\mu}H_{int}$ (Fig.5), with
$\gamma_{\mu}$ the muon gyromagnetic ratio and $H_{int}$ the local
magnetic field at the muon generated by the collinear order. In a
powder about 2/3 of the total signal oscillates while about 1/3 of
the muons are in a longitudinal field configuration. Hence, one
can write for the decay of the muon asymmetry
\begin{eqnarray}
 \nonumber
 A(t)= A_T\cdot {\rm cos}\, (\gamma_\mu H_{int} t +
         \phi_0) \cdot {\rm exp}\,(-\sigma t) \\
  + A_L\cdot {\rm exp}\,(-\lambda t) + A_{back}
  \,\,\,\,\,\,\,\,\,\,\,\,\,\,\,\,\,\,\,\,\,\,\,\,\,\,\,\,\,\,\,\,\,\,
  \label{muons2}
\end{eqnarray}
with $A_T$ the asymmetry of the oscillating component and $A_L$
the one of the longitudinal component. The ratio $A_T/A_L$
deviates from $2$ and is T-dependent since at a pulsed muon source
the amplitude of the oscillating part is progressively filtered
out as the precessional frequency increases. The reduction in
$A_T$ on cooling is actually slightly more pronounced than what
one would expect, taking into account that at ISIS the width of
the muon pulse is about 70 ns. This suggests that a small part (a
fraction below 15\% ) of the initial asymmetry is lost due to fast
relaxing muons. This observation is supported by preliminary
experiments performed at PSI facility. There it was found that the
asymmetry decay is characterized by a certain distribution of
relaxation rates, typical of systems with impurities, and that a
fraction of muons relax too fast to be detected at ISIS pulsed
muon facility. In Fig.\,\ref{field} the T-dependence of $H_{int}$
for $x=0.05$ is reported and compared to the one for $x=0$
\cite{musr}. One observes only a slight reduction of the
saturation value of the local field at the muon and of $T_c$,
while the temperature dependence is unaffected.

\begin{figure}[t!]
\vspace{10cm} \includegraphics{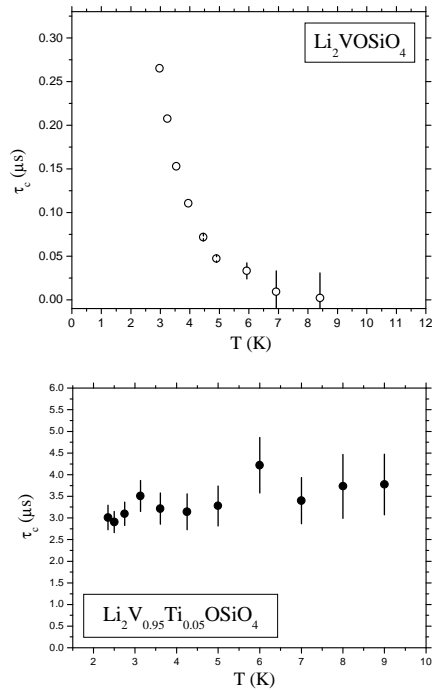}  \caption{Temperature dependence of the
correlation time for the low frequency fluctuations in
Li$_2$VOSiO$_4$ (top) and Li$_2$V$_{1-x}$OTi$_x$SiO$_4$ for
$x=0.05$ (bottom).}\label{tauC}
\end{figure}

Above $T_c$ the decay of the muon is conveniently described by a
nearly static Kubo-Toyabe function. As in the pure compound the
decay of the muon polarization can be conveniently fit using Keren
analytical approximation \cite{keren}, which yields reliable
results in the fast fluctuating limit, namely when
$\tau_c\gamma_{\mu}\sqrt{<\Delta h^2>} <<1$, with $\tau_c$ the
characteristic correlation time for the modulation of a magnetic
field distribution of amplitude $\sqrt{<\Delta h^2>}$. At variance
with \lisi \, in the Ti doped samples one finds that
$\tau_c\gamma\sqrt{<\Delta h^2>} \simeq 1$. However, Keren
approximation can still be used provided the fit is limited to a
time $t\simeq \tau_c$ \cite{keren}. In fact, the data above $T_c$
were accurately fit up to $t=6$ $\mu s$ with the function (Fig.5)
\begin{equation}
A(t) = A(0)exp(-\lambda t) P^{K}(H,\tau_c,<\Delta h^2>) ,
\end{equation}
where the first term describes the spin-lattice relaxation driven
by the fast fluctuations, while the second one is Keren analytical
approximation of Kubo-Toyabe function. By fixing $\lambda$ from
the high longitudinal field measurements above $900$ Gauss and
fitting $\mu$SR data at different magnetic fields, the
T-dependence of $\tau_c$ was derived. It was found that $\tau_c$,
at variance with the pure compound where it diverges on cooling,
is $T-$independent and more than an order of magnitude larger
(Fig.\,\ref{tauC}).

\section{Discussion}

In the weak doping limit, when the so called ``dilution model"
should hold, the spin Hamiltonian of a 2DQHAF with only n.n.
interactions  can be written in the form
\begin{equation}
\label{eq:2} {\cal{H}}=J\,\sum_{i,j}\,p_i\,p_j\,{\bf S}_i\cdot{\bf
S}_j=J(0)\,(1-x)^2\,\sum_{i,j}\,{\bf S}_i\cdot{\bf S}_j ,
\end{equation}
where the $x=0$ Hamiltonian is modified simply by taking into
account the probability $p_i$ to find a spin at site $i$. This
leads to a simple renormalization of the characteristic energy
scales and the spin stiffness becomes {\cite{uno}}
$\rho_s(x)=\rho_s(0) \cdot (1-x)^2$. One can try to extend this
result to frustrated 2DQHAF on a square lattice by considering
that also a second-nearest neighbour coupling is present. On
qualitative grounds one would expect that, since both the
probability to find a pair of nearest-neighbour spins and of
next-nearest-neighbours scale as $(1-x)^2$, either $J_1$ or $J_2$
are renormalized by the same factor and hence, the effective
degree of frustration $J_2/J_1$ is unaffected by dilution.
Moreover, the spin-stiffness is expected to scale as $(1-x)^2$ as
for the non-frustrated system. It will be shown hereafter that
although the spin-stiffness scales roughly as $(1-x)^2$ the
experimental data evidence a modification of the effective
$J_2/J_1$ ratio upon increasing $x$.

Preliminary information on the effects of spin-dilution can be
obtained from the analysis of the doping dependence of the
transition temperature $T_c(x)$, which shows a different trend
with respect to non-frustrated 2DQHAF.  In Fig.\,\ref{fig:4}
$T_c(x)$, extracted either from susceptibility data or from the
peak in $^7$Li  $1/T_1$, is shown. The initial suppression rate of
$T_c$ with Ti doping $-d\,T_c(x)/d\,x=C\,T_c(0)$, is not
characterized by a $C$ value close to the one theoretically
predicted and experimentally found for non-frustrated 2DQHAF,
namely $C\simeq 3.2$ \cite{McGurn}. In fact, in \liti \, $T_c(x)$
is rather well described by the expression $T_c(x)= T_c(0) (1-2x)$
(see Fig. \ref{fig:4}), namely by $C\simeq 2$. This difference can
be explained by resorting to the mean-field expression for
$T_c(x)$ \cite{Neto}

\begin{equation}
\label{CastroNeto} k_B\,T_c(x)= J_{\perp} (1-x)^2 \xi(x,T_c)^2
\left(\frac{M(x)}{M(0)} \right)^2
\end{equation}
where $M(x)$ is the $T=0$ staggered magnetization, $\xi(x,T)\sim
exp(2\pi\rho_s(x)/T)$ the in-plane correlation length in lattice
units \cite{CHN} and $J_\perp$ the inter-layer coupling, which is
reduced by a factor $(1-x)^2$ accounting for the probability to
find two coupled spins in adjacent layers.

From this expression it is evident that the reduction of the
staggered magnetization induced by dilution contributes to the
reduction of $T_c(x)$. However, at variance with non-frustrated
compounds, in which a spin dilution of 5\% was found to reduce the
zero temperature magnetization already by about 9 \% \cite{Corti},
in {\liti} the same amount of doping reduces it only by $1 \%$
(see Fig. \ref{field}). This explains why in \liti \, the effect
of doping on $T_c(x)$ is less pronounced. Moreover, it is observed
that if in Eq.\ref{CastroNeto} one neglects the reduction of the
staggered magnetization with doping and one considers that
$\rho_s(x)\sim \rho_s(0)(1-x)^2$, one finds an initial suppression
of $T_c(x)$ characterized by a coefficient $C=2$, exactly the one
experimentally observed for \liti . In the lower part of Fig.2 the
$x$-dependence of the spin-stiffness, estimated from Eq.5 using
the experimental values for $T_c(x)$ and neglecting the reduction
with $x$ of the staggered magnetization, is reported. It is
observed that the spin-stiffness is reduced by a factor $\simeq
(1-x)^2$ by spin dilution. A more accurate estimate of $\rho_s(x)$
can be done taking into account the decrease of $M(x)/M(0)$ in
Eq.5, derived from the reduction of the local field at the muon
for $T\rightarrow 0$ (Fig.6). One finds that
$\rho_s(x)/\rho_s(0)\simeq 1- 1.5\, x$. Hence one concludes that
the reduction of the spin-stiffness with $x$ is close to but not
exactly the one that one would obtain by rescaling $J_1$ and $J_2$
by exactly the same factor $(1-x)^2$.

In this respect it is interesting to analyze the effect of
Ti-doping on the effective ratio between the competing exchange
couplings. A first evidence that $J_2(x)/J_1(x)$ is not
$x-$independent comes from the analysis of the susceptibility data
(see Fig.\,\ref{fig:4}). The ratio between $T^m_{\chi}$ and the
Curie-Weiss temperature $\Theta$ is a measure of the degree of
frustration \cite{sei}. In fact, while for $J_2=0$ this ratio is
close to unity, on increasing $J_2$ it diminishes, reaching a
minimum for $J_2/J_1=0.5$ and then increases again \cite{sei}. In
{\liti} this ratio is found to slightly decrease with $x$ from
$T^m_{\chi}/\Theta=0.57 \pm 0.02$ for $x=0$ to
$T^m_{\chi}/\Theta=0.48 \pm 0.04$ for $x=0.11$. In principle it is
difficult to extract precise values of $J_2/J_1$ from the above
ratios, however, if one considers that for small changes $J_2/J_1$
varies linearly with $T^m_{\chi}/\Theta$, the observed
modifications imply a reduction in $J_2/J_1$ by 15-20 \% for
$x=0.11$.

Also the analysis of the temperature dependence of the nuclear
spin-lattice relaxation rate $1/T_1$ suggests that the effective
ratio $J_2/J_1$ is affected by spin-dilution. $1/T_1$ can be
written in terms of the components of the dynamical structure
factor $S({\bf q},\omega)$ at the nuclear Larmor frequency
$\omega_L$ as
\begin{equation}
\label{eq:1} \frac{1}{T_1}=\frac{\gamma^2}{2N}\sum_{\bf q}|A_{\bf
q}|^2\,S({\bf q},\omega_L)
\end{equation}
where $\gamma$ is the nuclear gyromagnetic ratio and $|A_{\bf
q}|^2$ is the form factor, which describes the hyperfine coupling
of the spin excitations at wavevector {\bf q} with the nuclei.

In non-frustrated two-dimensional $S=1/2$ antiferromagnets it was
pointed out  that the in-plane correlation length can be
conveniently derived from the nuclear spin-lattice relaxation rate
\cite{Sala} . In fact, by applying scaling arguments for the
amplitude and frequency of the collective spin excitations at
wavevector {\bf q}, one can write the NMR relaxation rate in terms
of the correlation length $\xi$ as \cite{Rep,Blinc}

\begin{equation}
\label{eq:3}
 \frac{1}{T_1}=
 C\,\varepsilon\,
 \xi^{z+2}\,\frac{\beta^2\,\sqrt{2\pi}}{\omega_E}\,(\frac{1}{4\pi^2})
 \int_{BZ} d{\bf q}\,\frac{|A_{\bf q}|^2}{(1+q^2\xi^2)^2}
\end{equation}
where $C=\gamma^2\,S(S+1)/3$, $z$ is the dynamical scaling
exponent (for $J_2/J_1=0$ $z\simeq1$ \cite{Sala}), $\varepsilon$
is a coefficient accounting for the reduction of the amplitude of
spin excitations due to quantum fluctuations , $\omega_E\sim
\sqrt{J_1^2+J_2^2}k_B/\hbar$ is the Heisenberg frequency
\cite{Melzi2}, describing the uncorrelated spin fluctuation for $T
\rightarrow \infty$ and $\beta$ is a normalization factor which
preserves the spin sum rule. The form factors have been calculated
on the basis of the hyperfine constants determined for the pure
\lisi \, \cite{Melzi1,Melzi2}.

\begin{figure}[t!]
\vspace{5.4cm} \includegraphics{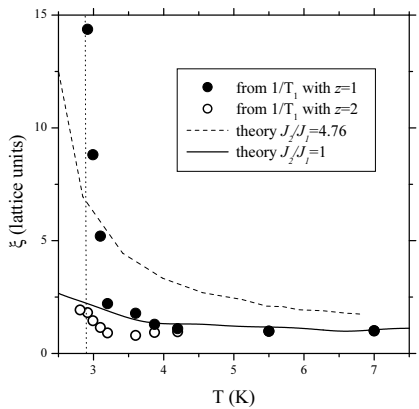} \caption{Correlation length $\xi$
extracted from $^7$Li NMR $1/T_1$ in Li$_2$VOSiO$_4$ by inverting
Eq.\,(\ref{eq:3}) in the text, in the assumptions that $z=1$ or
that $z=2$. The data are compared with theoretical calculations,
for different values of $J_2/J_1$, made in the framework of
pure-quantum self-consistent harmonic approximation
\cite{capriotti}. The vertical dotted line indicates the
transition to the collinear phase. }\label{capriotti}
\end{figure}

Once $z$ is defined a one-to-one relationship between $T_1$ and
$\xi$ is established, and one can determine $\xi$ from $^7$Li
$1/T_1$. Then one can compare these results for $\xi$ with the
recent theoretical predictions for the temperature dependence of
$\xi$, for different values of $J_2/J_1$, based on the
pure-quantum self-consistent harmonic approximation
\cite{capriotti}. In Fig.\,\ref{capriotti} $\xi$ extracted from
$^7$Li NMR relaxation data in Li$_2$VOSiO$_4$ are reported both in
the assumptions that $z=1$ or that $z=2$.

One observes that, above the crossover from 2D to 3D yielding the
divergence of $1/T_1$ at $T_c$, $\xi$ data derived for $z=1$ are
in rather good agreement with the theoretical calculations for
$J_2/J_1=1$. Larger ratios of $J_2/J_1$ and a dynamical scaling
exponent $z=2$ do not appear to be consistent with the
experimental findings. This suggests that the same scaling laws
used for pure 2DQHAF could still be valid in the presence of
frustration. For $z=1$ and $T\ll J_1+J_2$, if the form factor is
weakly $q$-dependent as for $^7$Li in \lisi ,  one has that
\cite{Sala,CHN}
\begin{equation}
\label{scaling}  \frac{1}{T_1}\,\sim\, \xi\, \sim \,{\rm
exp}\,(2\,\pi\,\rho_s(x)/T).
\end{equation}

\begin{figure}[t!]
\vspace{9cm} \includegraphics{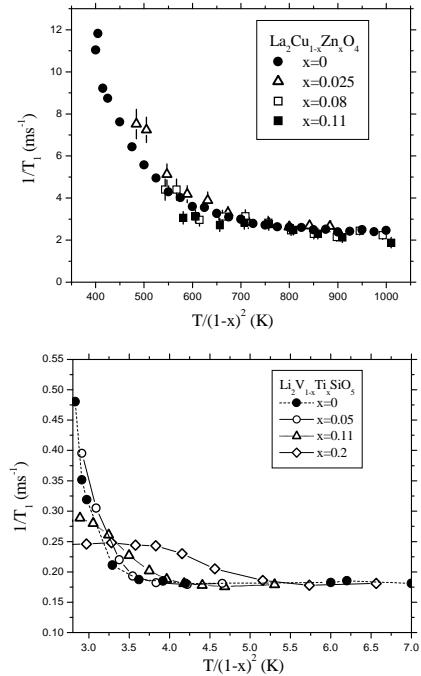}  \caption{Comparison of $^{63}$Cu $1/T_1$
in La$_2$Cu$_{1-x}$Zn$_x$0$_4$ and $^7$Li $1/T_1$ in
Li$_2$V$_{1-x}$OTi$_x$SiO$_4$ as a function of the scaled
temperature $T/(1-x)^2$.
 }\label{fig:5}
\end{figure}

Then, if $J_1$ and $J_2$ are reduced to the same extent by
dilution and $\rho_s(x)=\rho_s(0) \cdot (1-x)^2$, the T-dependence
of $1/T_1$ data should be $x$ independent once the temperature is
rescaled to $T/(1-x)^2$. In fact, in diluted non-frustrated 2DQHAF
as La$_2$Cu$_{1-x}$Zn$_x$O$_4$, such a scaling does occur (see
Fig.\,\ref{fig:5}). On the other hand, in the diluted frustrated
2DQHAF \liti, such scaling does not seem to be accurate any
longer, suggesting that also the effective ratio $J_2/J_1$ is
changing. In fact, one has to observe that while the spin
hamiltonian for a diluted non-frustrated 2DQHAF on a
square-lattice can still be mapped onto the same hamiltonian,
provided the exchange coupling is renormalized, the spin
hamiltonian for the frustrated 2DQHAF on a square lattice can no
longer be mapped onto the same hamiltonian as dilution in the
$J_2-J_1$ model yields the appearance of triangular spin
configurations.

After having presented these arguments suggesting  a variation of
the effective ratio $J_2(x)/J_1(x)$ and a negligible reduction of
$M(x)$, the effect of spin vacancies on the distortion initially
observed in undoped frustrated systems will be discussed
\cite{Melzi1}. As pointed out in the introduction, \lisi \, is
characterized by $J_2/J_1>0.8$ and hence by a two-fold degenerate
collinear ground-state, with a magnetic wave vector which can be
either {\bf Q}=(0, $\pi/a$) or {\bf Q}=($\pi/a$, 0). At low-T,
after an Ising transition, the spin system will eventually
collapse in either one of the two collinear ground-states
\cite{Chandra2,BeccaI}. Now, in a real system a finite
spin-lattice coupling exists and for $J_2/J_1>0.8$ it would favour
a tetragonal to orthorombic distortion \cite{Becca}. In fact, the
behaviour of $^{29}$Si NMR spectra and $1/T_1$ observed in some
\lisi \, samples is consistent with such a distortion
\cite{Melzi2}. $^{29}$Si form factor $|A_{\vec q}|^2$ is peaked at
(q$_x=\pi/a$, q$_y=\pi/a$) and vanishes at the critical wavevector
{\bf Q}=(0, $\pi/a$) or {\bf Q}=($\pi/a$, 0). Hence, on
approaching $T_c$, as the spectral weight shifts to the critical
wavevector of the envisaged {\it collinear } order, $1/T_1$ should
decrease and no peak observed. However, in the pure \lisi \, a
peak in $1/T_1$ was observed suggesting that a distortion
modifying $^{29}$Si form factor takes place \cite{Melzi1,Melzi2}.
In Ti-doped samples (Fig.4) no peak in $^{29}$Si $1/T_1$ is
observed at $T_c$, pointing out that Ti impurities tend to hinder
this distortion.

Finally, we point out that above $T_c$ Ti doping also hinders the
very-low-frequency dynamics evidenced  by $\mu$SR measurements in
{\lisi} (Fig. \ref{tauC}) \cite{musr}. These dynamics were
associated with the motion of domain walls separating regions were
correlations with {\bf Q}=(0, $\pi/a$) or {\bf Q}=($\pi/a$, 0)
develop before the lattice distortion  removes their degeneracy.
Although a clear explanation of such a phenomenon goes beyond the
aim of the present work, the observation of a T-independent much
longer correlation time measured by $\mu$SR suggests that Ti
impurities might pin the motions of domain walls.

\section{CONCLUSIONS}

In conclusion, a throughout investigation of a spin diluted
frustrated 2DQHAF system by means of NMR, $\mu$SR and
susceptibility measurements has been presented. It has been shown
that the decrease of the magnetic ordering temperature is
consistent with a reduction of the spin-stiffness by a factor
$\simeq (1-x)^2$ and with a minor effect of spin dilution on the
sublattice magnetization. The analysis of the magnetic
susceptibility and of the nuclear spin-lattice relaxation rate
shows that the effective ratio $J_2(x)/J_1(x)$ decreases with $x$,
namely that the two coupling constants are not renormalized in the
same way by dilution. This would actually mean that the spin
Hamiltonian of a frustrated 2DQHAF on a square-lattice cannot be
mapped onto the same square-lattice Hamiltonian after spin
dilution has occurred. Finally it was shown that the low frequency
dynamics observed in pure Li$_2$VOSiO$_4$ by means of $\mu$SR
measurements disappears in \liti . Moreover the absence of a peak
in $^{29}$Si NMR relaxation rate at the transition indicates that
the distortion induced by frustration might be hindered by doping.

\section*{Acknowledgements}

We would like to thank A. Fubini for sending us the results of the
theoretical calculations reported in Ref.\onlinecite{capriotti}
and T.Roscilde for useful discussions. This work was supported by
PRIN2004 National Project "Strongly Correlated Electron Systems
with Competing Interactions ".




\begin{references}
\bibitem{Tommaso} While for a two-dimensional square-lattice the
percolation threshold for site-dilution is $x\simeq 0.41$,
preliminary estimates carried out by T. Roscilde show that in
presence of next-nearest neighbour couplings along the diagonals
the percolation threshold increases to $x\simeq 0.58$.
\end{references}
\end{document}